\documentclass[useAMS,usenatbib]{mn2e} 
\usepackage{aas_macros}
\usepackage{graphics}
\usepackage[pdftex]{graphicx}
\usepackage{epstopdf}
\usepackage{epsfig}  
\usepackage{natbib} 
\usepackage{dingbat}
\usepackage{float}
\usepackage{amsmath}
\usepackage{times}
\usepackage[varg]{txfonts}
\usepackage{verbatim} 
\usepackage{multirow,bigdelim} 
\usepackage{color}
\usepackage{vmargin}
\setmarginsrb{1.2cm}{1cm}{1.2cm}{1cm}{1cm}{1cm}{1cm}{1cm}

\newcommand{\hmsun}{{\, h^{-1}\rm M}_\odot}

  \makeatletter
    \renewcommand{\paragraph}{\@startsection{paragraph}{4}{\z@}%
      {-3.25ex\@plus -1ex \@minus -.2ex}%
      {1.5ex \@plus .2ex}%
      {\normalfont\small\centering}}
     
    \renewcommand{\subparagraph}{\@startsection{subparagraph}{5}{\z@}%
      {-3.25ex\@plus -1ex \@minus -.2ex}%
      {1.5ex \@plus .2ex}%
      {\normalfont\small\centering}}
    \makeatother

\newcommand{\gadget}{{\sc Gadget}}

\newcommand{\music}{{\sc Music}}
\newcommand{\ramses}{{\sc Ramses}}

\newcommand{\kms}{{ km~s$^{-1}$}}
\newcommand{\hMpc}{{ \textit{h}$^{-1}$~Mpc}}

\title[Virgo Cluster]{Virgo: an unlikely cluster of galaxies because of its environment}
\author[Sorce et al.]
{{Jenny G. Sorce$^{1,2}$\thanks{E-mail: \text{jenny.sorce@univ-lyon1.fr / jsorce@aip.de}}, J\'er\'emy Blaizot$^{1}$, Yohan Dubois$^{3}$}\\
$^1$Univ Lyon, Univ Lyon1, Ens de Lyon, CNRS, Centre de Recherche Astrophysique de Lyon UMR5574, F-69230, Saint-Genis-Laval, France\\
$^2$Leibniz-Institut f\"{u}r Astrophysik, An der Sternwarte 16, 14482 Potsdam, Germany\\
$^3$Institut d'Astrophysique de Paris, UMR 7095 CNRS et Universit\'e Pierre et Marie Curie, 98bis Bd Arago, F-75014, Paris, France
}

\begin{document}

\date{}

\pagerange{\pageref{firstpage}--\pageref{lastpage}} \pubyear{2019}

\maketitle

\label{firstpage}

\begin{abstract}
\indent 
Galaxy clusters constitute powerful cosmological probes thanks to comparisons between observed and simulated clusters. As such Virgo constitutes a formidable source for detailed observations facilitated by its proximity.  However, the diversity of clusters complicates the comparisons on a one-to-one basis. Simulated clusters must be carefully selected, a daunting task since most properties are unknown. Alternatively, lookalikes produced in the proper large scale environment can be used. Additionally, their statistical study give access to the mean properties of the observed cluster including its most probable history as well as its deviation from an average cluster. This paper presents such a statistical study with 200 Virgo-like and 400+ cluster-size random dark matter halos. Only 18\%(0.5\%) of these random halos comply within 3(2)$\sigma$ with the mean values (radius, velocity dispersion, number of substructures, spin, velocity, concentration, center of mass offset with respect to the spherical center) of Virgo halos at z=0 and abide by a similar merging history up to redshift 4. None are within 1$\sigma$ because of environmentally induced properties (number of substructures and velocity). For further comparisons, random halos are selected to reproduce the mass distribution of the lookalikes to cancel mass bias effects. Redshift 1 appears then as a turning point: random to Virgo-like property ratios are alternatively smaller/larger than 1.  This highlights the importance of studying clusters within their proper large scale environment: simulated galaxy population, grandly affected by the cluster history, can then be compared with the observed one in details. Direct lookalikes simplify grandly the challenge.

\end{abstract}


\section{Introduction}
While galaxy clusters, as powerful cosmological probes, are extensively studied both observationally and numerically, comparing the observed and simulated clusters in detail can be quite challenging. The diversity of clusters in terms of morphologies, formation history, etc \citep{1988S&T....75...16S} indeed makes one-to-one comparisons a daunting task. The parameters that a numerical cluster should reproduce to be considered as an accurate lookalike of a given observed cluster are simply difficult to completely determine because of various aspects: lack of accuracy, no technique, weak knowledge of the correlation between parameters. In order to take up the challenge, we propose to focus on our nearest cluster neighbor: the Virgo cluster of galaxies. Because of its proximity,  this cluster has been and is still studied through numerous observational projects \citep[e.g.][for a non-exhaustive list]{2000eaa..bookE1822B,2009eimw.confE..66W,2011MNRAS.416.1996R,2012A&A...543A..33V,2011MNRAS.416.1983R,2011ASPC..446...77F,2012ApJS..200....4F,2012MNRAS.423..787T,2014MNRAS.442.2826C,2014ApJ...782....4K,2014A&A...570A..69B,2015A&A...573A.129P,2016ApJ...823...73L,2016ApJ...824...10F,2016A&A...585A...2B}.\\

The novelty of the present paper is to conduct a numerical statistical study of the properties of the Virgo cluster via a sample of unique Virgo candidates from constrained simulations of the local Universe \citep[e.g.][]{1987ApJ...323L.103B,2010arXiv1005.2687G,2010MNRAS.406.1007L,2013MNRAS.435.2065H,2016ApJ...831..164W}. These simulations differ from typical simulations \citep[e.g.][]{2012MNRAS.426.2046A,DeusSimulation2012,2016MNRAS.463.3948D} in the sense that they stem from initial conditions that have been constrained with local observational data. In our case, these observational data are radial peculiar velocities \citep[e.g.][]{1992ApJS...81..413M,1997ApJS..109..333W,2001MNRAS.326..375Z,2007ApJS..172..599S,2008ApJ...676..184T,2013AJ....146...86T,2016AJ....152...50T} but they can also be densities obtained with redshift surveys \citep[e.g.][]{2006AJ....131.1163S,aih11,2011MNRAS.416.2840L,2012ApJS..199...26H}. The constrained initial conditions for these simulations are constructed with different techniques either forwards \citep[e.g.][]{2008MNRAS.389..497K,2013MNRAS.432..894J,2013MNRAS.429L..84K,2014ApJ...794...94W} or backwards \citep[e.g.][]{1990ApJ...364..349D,1991ApJ...380L...5H,1992ApJ...384..448H,1993ApJ...415L...5G,1998ApJ...492..439B,1999ApJ...520..413Z,2008MNRAS.383.1292L,2016MNRAS.457..172L}. As a result, the simulations resemble the local Universe within a hundred megaparsecs down to a few megaparsecs \citep[e.g.][]{2016MNRAS.455.2078S}. \\

Based on a backwards scheme, described hereafter, applied to peculiar velocities, our simulations always host a unique dark matter halo,  lookalike of the Virgo cluster (position, mass, velocity, etc), in a large-scale environment that reproduces the local environment \citep{2016MNRAS.460.2015S}. Unlike numerical studies of clusters based on random simulations \citep[e.g.][]{2012MNRAS.420.2859M,2015ApJ...807...88G,2016MNRAS.tmp.1592L,2016MNRAS.457.4063S,2017MNRAS.tmp..205H,2017MNRAS.467.3827P,2017arXiv170305682D,2017arXiv170310907B}, this work allows studying statistically a given cluster, here Virgo, in its proper large scale environment. As a result of being in their proper environment, these dark matter halos present the same quiet merging history within the past 7 gigayears \citep{2016MNRAS.460.2015S} with on average only one merger, larger than about a tenth of their mass at redshift zero, within the last 4 gigayears \citep{2018A&A...614A.102O}. This last simulation-based finding has recently been supported by an observational study \citep{2018arXiv180804751L}. Since galaxy populations are not only sensitive to the large scale environment of the cluster \citep{2014A&A...562A..87E} but also to its formation history, in particular its past mergers \citep{2017arXiv170703208D}, such properties are fundamental requisites to legitimize comparisons between observed and simulated galaxy populations down to the details \citep{{2015ApJ...807...88G}}.\\

This paper starts with a description of the new large sample of 200 constrained cosmological simulations run to obtain the Virgo-like sample of dark matter halos at the core of this study. Random runs are also required to give a twofold goal to this paper. They supply random halos within the same mass range as the Virgo-like halos to be compared with. Subsequently, the third section explores the distribution of various properties (such as velocity, spin, number of substructures, etc) of the 200 Virgo-like halos as well as the evolution of these properties across cosmic time. The properties of the Virgo cluster are statistically determined. The fourth section compares random and Virgo halos. Properties of the Virgo cluster that are constrained and/or atypical on average are highlighted. The probability to find a Virgo-like cluster given the Planck standard cosmological model and the properties that discriminate the Virgo cluster from other clusters are determined.


\section{Simulations}

All the dark matter simulations are run with the adaptive mesh refinement code \ramses\ \citep{2002A&A...385..337T} within the Planck cosmology framework \citep[$\Omega_m$=0.307, $\Omega_\Lambda$=0.693, H$_0$=67.77\kms~Mpc$^{-1}$, $\sigma_8$~=~0.829,][]{2014A&A...571A..16P}. The resolution is set for halos to be constituted of a minimum of 10,000 particles at redshift zero.  Considering Virgo to be more massive than 10$^{14}$~$\hmsun$, it is equivalent to a particle mass of about 10$^9$~$\hmsun$. The initially coarse grids are adaptively refined down to 3.8~h$^{-1}$~kpc. Such a resolution allows us to probe also higher redshifts than zero. Simulations are run from redshift 120 to redshift zero. Note that the N-body code used in this paper is different from that used in our previous papers \cite[e.g][]{2016MNRAS.455.2078S} for two reasons: 1) results obtained with two different N-body code are supposed to be similar in the dark matter only case \citep{2016MNRAS.458.1096E}. We check that this is indeed the case for our constrained simulations ; 2) in a perspective of hydrodynamical simulations of the Virgo cluster, the adaptive mesh refinement grid code will be used. It is thus appropriate to use it to select the candidates from the sample of 200 halos that will be run with gas.

\subsection{Constrained simulations}

\begin{figure*}
\includegraphics[width=0.9\textwidth]{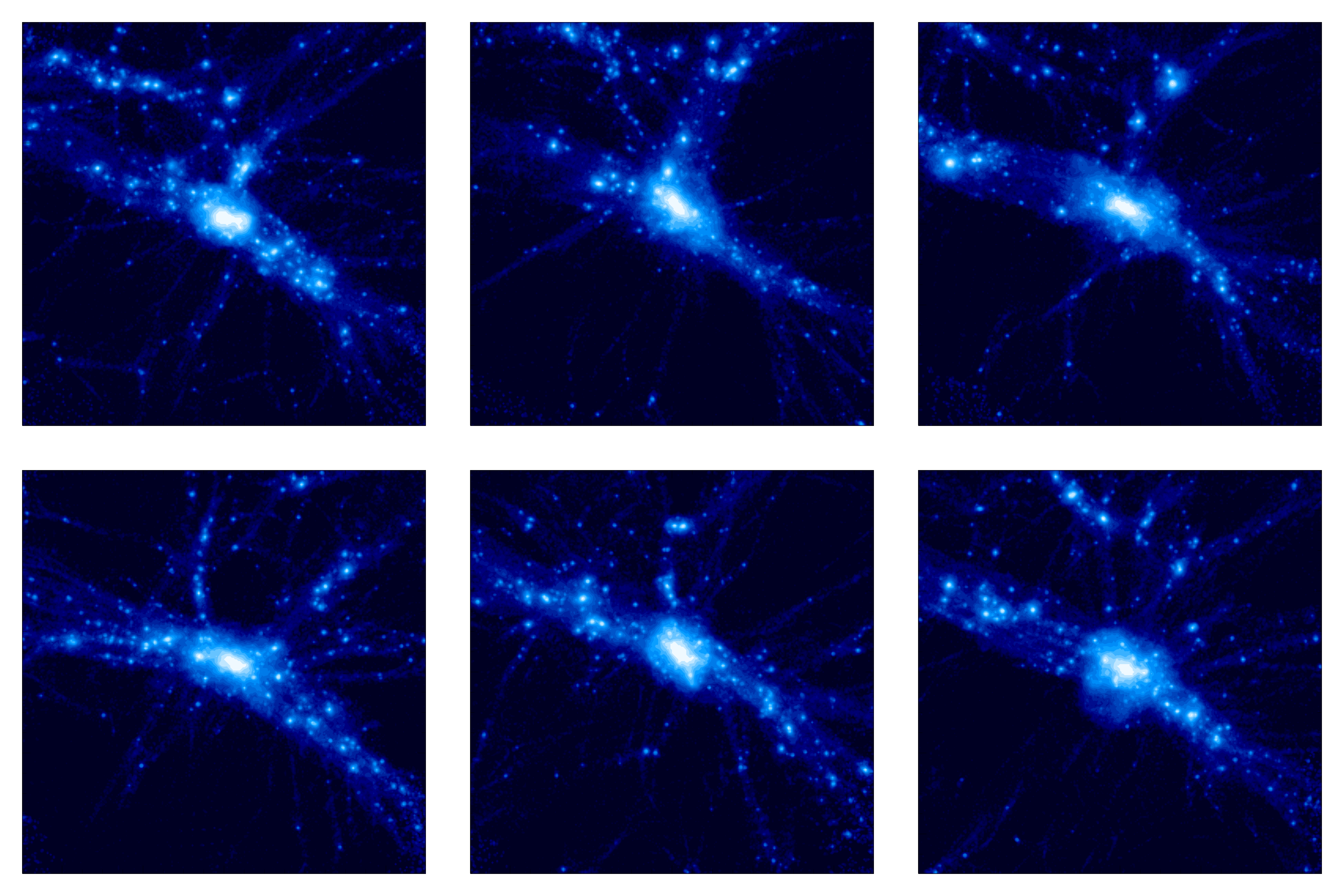}
\caption{20~\hMpc\ XY supergalactic slices of 5~\hMpc\ thickness representing 6 Virgos. The constrained boxes are oriented in the same way as the local Universe to obtained the X, Y and Z coordinates. The similarities (positions, filaments, orientation, etc) between the 6 Virgo halos is remarkable. Two Virgo halos have been selected because the values of the 8 properties mentioned in Figure \ref{fig:proba} match the entire set of Virgo halos' mean property values. One Virgo halo has been selected because its merging history matches the mean merging history of the Virgo set. The other three have been completely randomly extracted from the 200 Virgo-like halos sample.}
\label{fig:virgos}
\end{figure*}

Before running the simulations where the Virgo-like halos can be found, initial conditions, constrained by observational data (in our case galaxy peculiar velocities) for the resulting simulations to resemble the local Universe, must be prepared. The different steps of the whole scheme used to build this set of constrained initial conditions are described in detail in \citet{2016MNRAS.455.2078S}. 

A brief description of these steps and their purpose are reminded here:
\begin{itemize}
\item Grouping of the radial peculiar velocity catalog to remove non-linear virial motions \citep[e.g.][]{2015AJ....149..171T,2015AJ....149...54T} that would affect the linear reconstruction obtained with the linear method \citep[e.g.][]{2017MNRAS.468.1812S,2017arXiv170503020S}. Namely, when measurements are available for several galaxies in a given cluster, these measurements are replaced by one measurement, that for the cluster.
\item Minimizing the biases \citep{2015MNRAS.450.2644S} in the grouped radial peculiar velocity catalog. Biases are indeed inherent to any observational catalog and in this particular case give rise to a spurious overall infall onto the local volume.
\item Reconstructing the 3D cosmic displacement field with the Wiener-Filter technique \citep[linear minimum variance estimator, in abridged form WF,][]{1995ApJ...449..446Z,1999ApJ...520..413Z} applied to the radial peculiar velocity constraints.
\item Relocating constraints to the positions of their progenitors using the Reverse Zel'dovich Approximation and the reconstructed cosmic displacement field \citep{2013MNRAS.430..888D} and replacing noisy radial peculiar velocities by their 3D WF reconstruction \citep{2014MNRAS.437.3586S}. Subsequently, one can expect structures to be at the proper position, i.e. at positions similar to those observed, at the end of the simulation run.
\item Producing density fields constrained by the modified observational peculiar velocities combined with a random realization to restore statistically the ``missing'' structures. The WF indeed goes to the null field in absence of data or in presence of very noisy data. The Constrained Realization technique \citep[CR,][]{1991ApJ...380L...5H,1992ApJ...384..448H}, that differs schematically from the WF by a random realization added to the constraints, is used for that step. Note that this field is the first source of residual cosmic variance between the simulated Virgo halos.
\item Rescaling the density fields to build constrained initial conditions and eventually increasing the resolution by adding small scale features \citep[e.g. \music\ code,][]{2011MNRAS.415.2101H}. These small scale features are the second source of variance between the Virgo halos but only at the non-linear level. 
\end{itemize}

To avoid periodicity problems in the local Universe-like region, the boxsize for the 200 constrained simulations is set to 500~\hMpc\ at z=0 \citep{2016MNRAS.455.2078S}. To decrease the computational cost of the 200 runs and since our interest lays solely in the study of the Virgo cluster, the zoom-in technique, first proposed by \citet{2001ApJS..137....1B} and implemented in \music, is used. The zoom-in technique consists in keeping the large scale environment at low resolution for its effect on the region of interest while the resolution is increased solely in this region. This requires knowing the position of the particles of the Virgo progenitors and their surroundings in the initial conditions. 

We then proceed as follows to minimize the computational cost:
\begin{itemize}
\item the 200 constrained initial conditions are run at a low resolution (256$^3$ particles, i.e. a particle mass of 6$\times$10$^{11}$~$\hmsun$ or about 100 particles per Virgo-like halos at redshift zero). At this stage, we are only interested in getting the particles that constitute Virgo and its surroundings within a 10~\hMpc\ radius sphere to trace them back to the initial redshift in order to get their initial position. This resolution is then sufficient. 
\item The Virgo lookalike is identified in each one of these simulations using the list of halos obtained with Amiga's Halo Finder\footnote{Note that first, \ramses\ outputs are converted to \gadget\ format so that they can be processed by the halo finder.} \citep{2009ApJS..182..608K}. 
\item Particles within a 10~\hMpc\ radius sphere around the center of the candidates are traced back to the initial redshift. They define the zoom-in region given to \music\ to produce the initial conditions with a higher resolution in that region than in the rest of the box.
\item After running \music\ with an effective resolution of 2048$^3$ particles for the highest level (particle mass 1.2$\times$10$^9$~$\hmsun$), the zoom-in initial conditions are run with \ramses\ and analyzed with the Amiga's Halo Finder.
\item After building the merger tree of the Virgo-like halos, the evolution of their properties across cosmic time are stored.
\end{itemize}

\subsection{Random simulations}

For reference, we run 3 random initial conditions prepared with \music\ with 1024$^3$ particles within a full 250~\hMpc\ box. This boxsize and this number of particles ensure the same mass resolution (particle mass 1.2$\times$10$^9$~$\hmsun$) for both the random and the Virgo-like halos. 

In these 3 simulations, a total of about 400+ halos have a mass above 1.5$\times$10$^{14}$~$\hmsun$ (mass of the smallest Virgo-like halo among the 200 available) but below 10$^{15}$~$\hmsun$ \citep[reasonable mass upper limit for the Virgo cluster, e.g.][]{2011A&A...532A.104N,2010MNRAS.405.1075K,2014ApJ...782....4K,2015AJ....149..171T} using the `M$_{200}$' definition\footnote{Since results and conclusions drawn from the halo samples are identical using either the `M$_{200}$' definition or the virial definition, we choose to present only those obtained with the `M$_{200}$' definition.} (i.e the mass enclosed in a sphere with a mean density of 200 times the critical density of the Universe).  Hereafter, these random halos are referred to as cluster-size random halos.\\

This cluster-size random halo sample is however biased towards the low mass end when compared to that of the Virgo-like. To remove mass bias effects from the comparisons, an additional selection criterion is applied to the sample of cluster-size random halo. The random sample should follow a similar mass distribution as that of the Virgo halo, in other words we set the mass distribution, M$_{0}^{\rm s}$, of the random sample. To that end:\\
1) random halos are selected to have masses between [$\overline{\rm M_{0,\rm virgo}}$-n~$\times~\sigma_{\rm Mo,virgo}$ , $\overline{\rm M_{0,\rm virgo}}$+n~$\times~\sigma_{\rm Mo,virgo}$]  where n=2, $\sigma_{\rm Mo,virgo}$ is the standard deviation of the Virgo-like masses and $\overline{\rm M_{0,\rm virgo}}$ the average mass of the Virgo-like halos at redshift zero\footnote{Note that n=1 or n=3 give the same results.}. This selection permits removing extreme cases from the subsample.\\
2) the distribution is then forced to have the same mean and standard deviation as that of the Virgo halos at redshift zero. The low mass end of the subsample needs then to be resampled since there is an excess of low mass halos. Namely, only a few of the lower mass random halos are randomly selected among the random halos at	 the low mass end.  The random and constrained M$_0$ then match each other in terms of mean and standard deviation. The other moments (skewness and kurtosis) of the two distributions, in addition to be small, are found to be both of the same order of magnitude and same sign. Consequently, the distribution can be considered similar. Note that the few randomly selected halos among the random halos at the low mass end alter neither the results nor the conclusions. \\

This strategy allows us to get a mass-unbiased sample of random halos  while preserving a high enough number of halos for statistical purposes. Still such a selection drastically reduces the number of random halos available for comparisons: Only 77 halos (18\%) are left in the random sample, giving already an idea of the commonness of Virgo-like halos.

\section{Properties of the Virgo cluster with constrained simulations}

Figure \ref{fig:virgos} shows 20~\hMpc\  XY supergalactic slices with a 5~\hMpc\ thickness of 6 Virgo halos. Boxes are oriented in the same way as the local Universe to identify the X, Y and Z supergalactic coordinates. The gradient of colors stands for the dark matter density field. The visual similarities (positions, filaments and their orientations) between the different Virgo halos are already impressive compared to typical simulations of random clusters.\\

\begin{figure*}
\vspace{-1cm}
\hspace{-0.8cm}\includegraphics[width=1\textwidth]{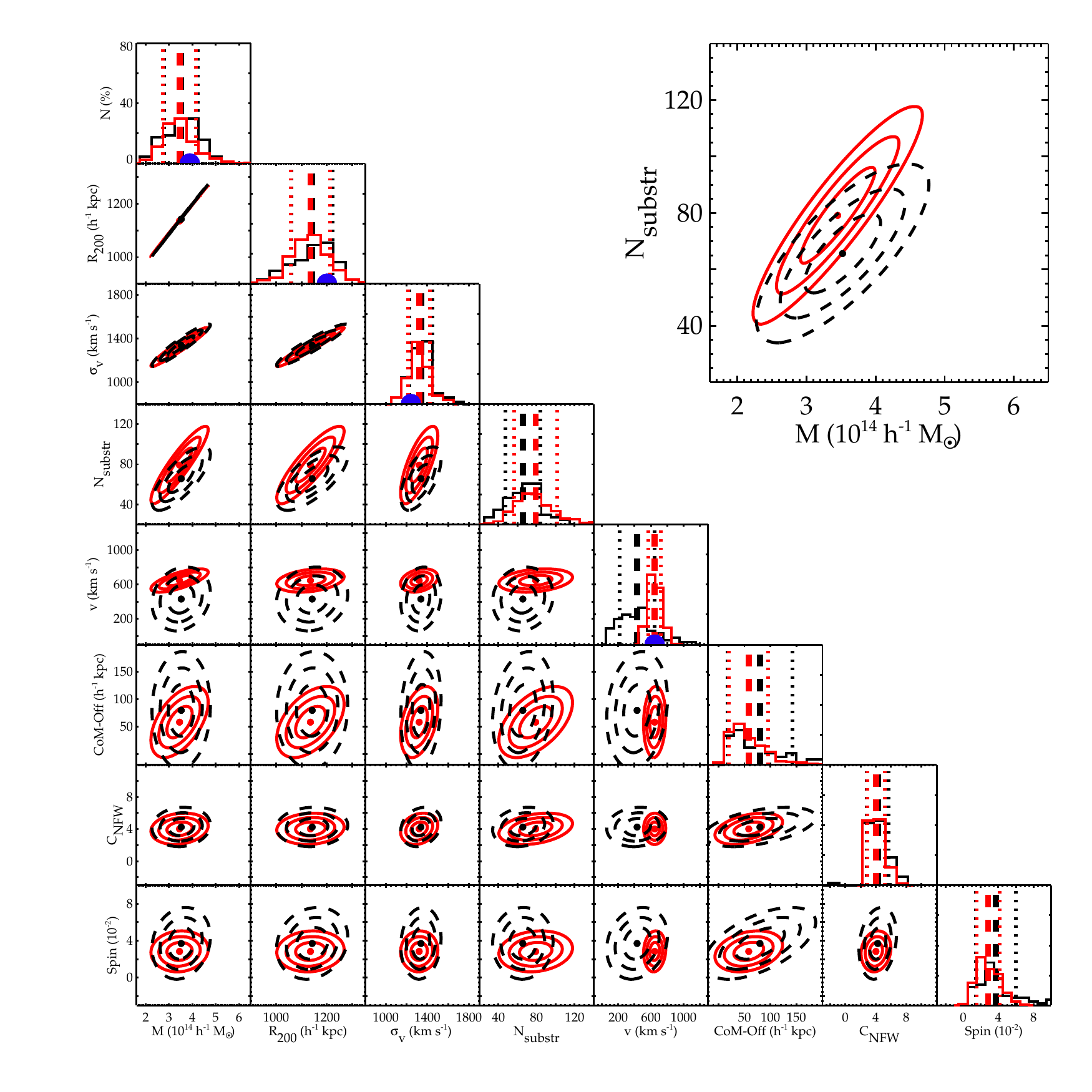}
\vspace{-0.5cm}
\caption{Normalized and joint property distributions of the mass-unbiased random halo sample (black) and Virgo halo sample (red). Thick dashed and thin dotted lines stand for the mean and standard deviation of the normalized distribution using the same respective colors. The filled circles and the ensemble of three ellipses are the mean and 1, 2 and 3-$\sigma$ probabilities of the joint distributions. Random halos have masses between  [$\overline{\rm M_{0,\rm virgo}}$-2~$\times~\sigma_{\rm Mo,virgo}$ , $\overline{\rm M_{0,\rm virgo}}$+2~$\times~\sigma_{\rm Mo,virgo}$]  where $\sigma_{\rm Mo,virgo}$ is the standard deviation of the Virgo-like masses and $\overline{M_{0,\rm virgo}}$ is the mean mass of the Virgo-like halos at z=0. The blue filled circles on the normalized distributions are observational or reconstructed estimates of the properties of the observed Virgo cluster. From left to right or top to bottom, properties are mass, radius, velocity dispersion, number of substructures, velocity, center of mass offset with respect to the spherical center, concentration and spin}
\label{fig:proba}
\end{figure*}

Figure \ref{fig:proba} gives the joint distributions as well as the normalized distributions of the parameters of the Virgo halos at redshift zero in red. The black color stands for the random halos. Joint distributions are obtained with covariance matrices. Given that the third and fourth moments of the random and constrained parameter distributions are of the same order of magnitude and of the same sign, the use of covariance matrices is legitimate in a first approximation for simple visual comparisons. This figure is further analyzed for the sole Virgo cluster in the following in two steps: 1) observed versus simulated properties, 2) statistically derived properties of the Virgo cluster from the constrained simulations.

\subsection{Observed vs. Simulated Virgos: general agreement}

Mean and standard deviation (1$\sigma$) of the normalized distributions are indicated respectively by red dashed and dotted lines. The blue filled circles on these normalized distributions stand for observational/reconstructed estimates of the properties of the Virgo cluster. The mass and the radius are from \citet{2017arXiv171008935S}. The mass is obtained with a first turn around radius study and is thus converted to M$_{200}$ using a 0.7 factor \citep{2016MNRAS.460.2015S} and H$_0$=67.77~\kms~Mpc$^{-1}$ (Planck cosmology). The radius is that of collapsed matter (roughly equivalent to the virial radius). It is converted to R$_{200}$ using a 0.8 factor (obtained when comparing R$_{200}$ and R$_{\rm vir}$ derived by the Halo finder and assuming that observational and numerical definitions of the virial radius give similar results).
The line-of-sight velocity dispersion, dispersion of the velocity of the galaxies belonging to the cluster around their mean velocity projected along the line-of-sight direction, \citep{2015AJ....149..171T} is multiplied by $\sqrt{3}$ assuming a fairly isotropic dispersion. The velocity of the cluster with respect to the CMB, hereafter velocity, is reconstructed with the Wiener Filter \citep{2015MNRAS.450.2644S}. Other property observational estimates are not available except perhaps the number of substructures. However this would require a more detailed study especially to retain only the substructures of a given mass. That would permit also further studies including the position, merging history, population of the substructures, etc \citep{2014A&A...570A..69B}. We thus postpone this study to another paper. \\

Regardless, the agreement between observational/reconstructed (blue) and simulated (thick red dashed line for the mean) estimates are remarkable given the numerous assumptions and the difficulty in comparing observed uncertain estimates and simulated properties.

\subsection{Simulated Virgos: statistically determined properties of the cluster}

\begin{table}
\begin{center} 
\begin{tabular}{lr@{ }rr}
\hline
\hline
 Parameter  &  \multicolumn{1}{c}{Mean} & Standard deviation \\
   \hline
M$_{200}$ ($\hmsun$) & 3.45e+14  & 7.1e+13  \\
R$_{200}$ ($h^{-1}$kpc) & 1135  & 77 	    \\
 $\sigma_v$ (km s$^{-1}$ ) &	 1321 & 	104 	         \\
N$_{\rm substr}$ (M$>$1e+10) & 79 & 22  			                       \\
 v (km s$^{-1}$) &  646    & 79 \\
 CoM-off ($h^{-1}$ kpc) & 58  &  38\\
 C$_{\rm NFW}$ & 4 & 1\\
  spin & 0.03 & 0.01\\
 \hline
\hline
\end{tabular}
\end{center}
\vspace{-0.25cm}
\caption{Statistically derived values of the properties of the Virgo cluster.}
\label{Tbl:2}
\end{table}

Furthermore, on Figure \ref{fig:proba}, each red filled circle and ensemble of three ellipses stand respectively for the mean and 1, 2 and 3-$\sigma$ probabilities of the joint distributions. The panel on the top right corner is a zoom of the first panel of the fourth row of the main plot.\\

Expected correlations appear clearly between the mass, the radius, the velocity dispersion and the number of substructures. Another less expected correlation is visible between the mass (or radius, number of substructures, velocity dispersion) and the offset with respect to the center of mass: the more massive the Virgo cluster is the larger the offset is. Other much weaker correlations between the NFW concentration \citep[R$_{\rm halo}$/r$_{\rm s}$ with R$_{\rm halo}$ the radius of a halo and r$_{\rm s}$ the break radius between an inner r$^{-1}$ density profile and an outer r$^{-3}$ profile, see][]{2012MNRAS.423.3018P} and the number of substructures or the offset ; between the spin and the number of substructures or the offset seem to arise. Finally, Table \ref{Tbl:2} summarizes the statistically derived values of the properties of the Virgo cluster.

\section{Specificities and probability of the Virgo cluster}

\subsection{Virgo-like vs. Random Halos}

In Figure \ref{fig:proba}, the black color stands for the property distributions of the random halo sample with M$_0^{\rm s}$. By construction the mass distributions of random and constrained halos are very similar in the first row. Some obvious differences appear between the random and the constrained normalized and joint property distributions. The most interesting differences appear for the number of substructures, the offset with respect to the center of mass, the velocity and to a smaller extent the spin.\\

Before any further quantitative comparisons, we define two parameters relevant to determine whether a property is (a)typical or/and (not) constrained. A property is atypical if the mean value of the property for the random sample is significantly different from that obtained for the constrained sample. A property is constrained if the standard deviation of the property is larger in the random case than in the constrained case. Note that theoretically if the property is atypical, it can also be considered as constrained but for the sake of clarity, these two concepts are distinguished in the rest of the paper. It can happen that both conditions are fulfilled then the property is both constrained and atypical. To measure these conditions quantitatively, we use the formula given in equation \ref{eq:1} (the larger in absolute value $\Delta$ is, the more significantly different the means are) and the ratio of the constrained (virgo) and random (rand) standard deviations (the smaller with respect to 1 the ratio is, the more constrained the studied parameter is) respectively.

\begin{figure*}
\includegraphics[width=0.18\textwidth]{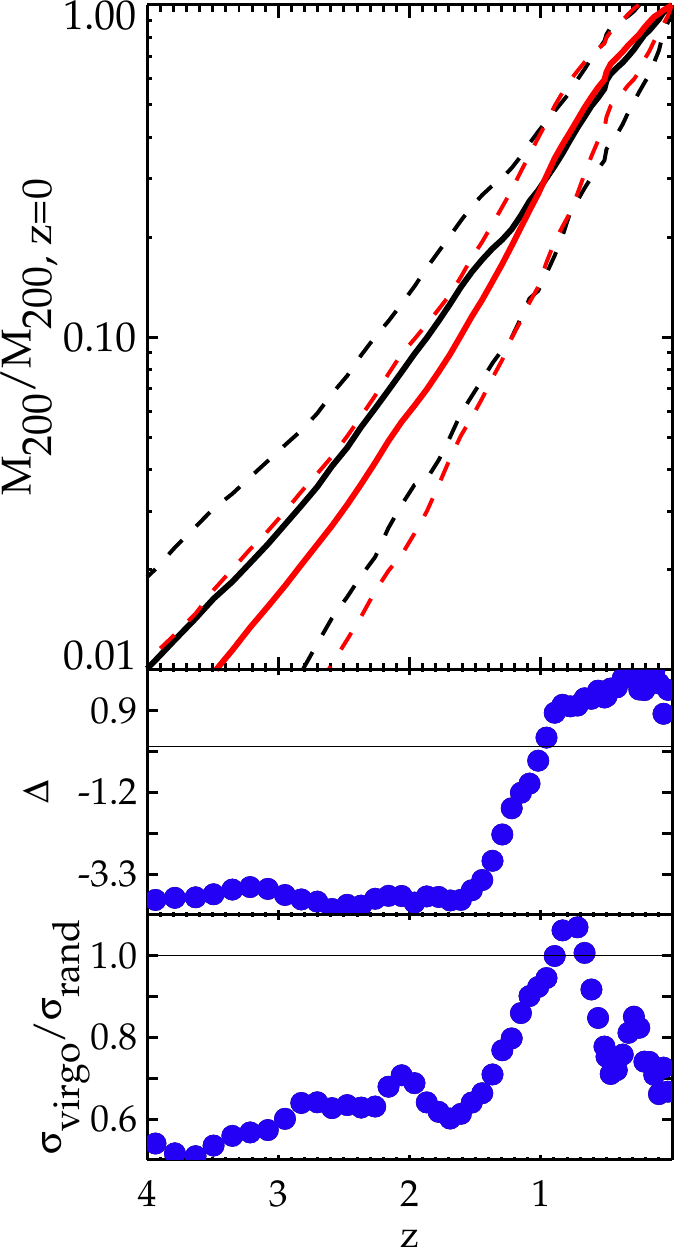}\hspace{0.5cm}
\includegraphics[width=0.179\textwidth]{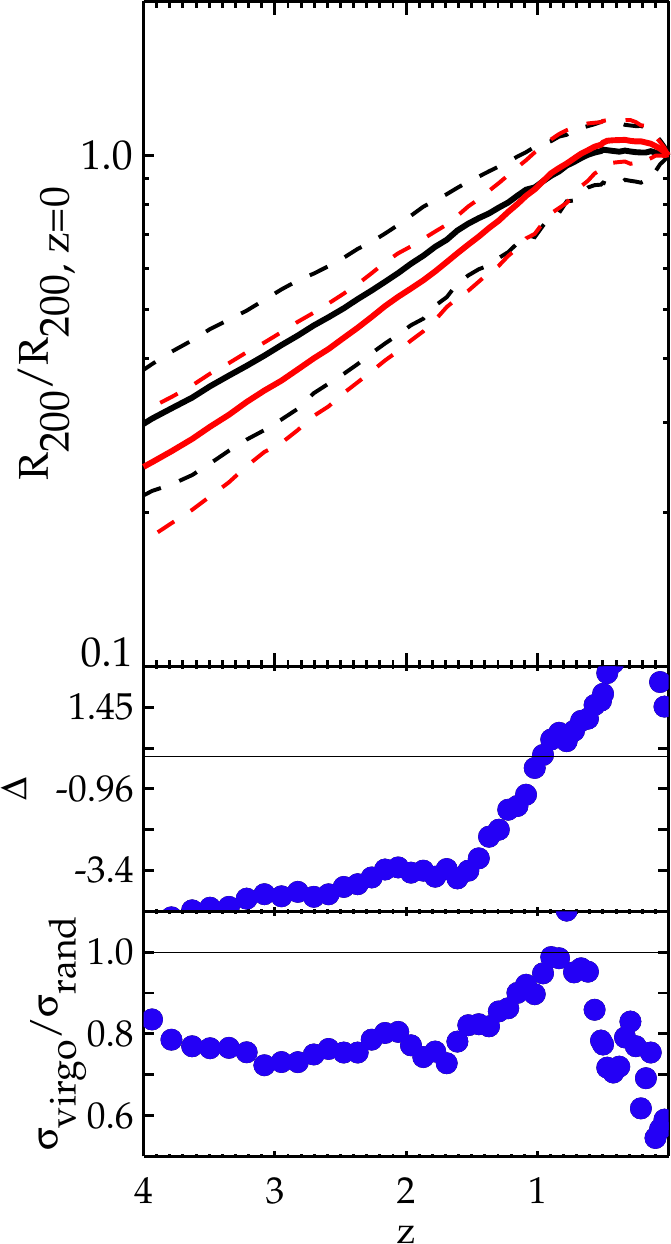}\hspace{0.5cm}
\includegraphics[width=0.182\textwidth]{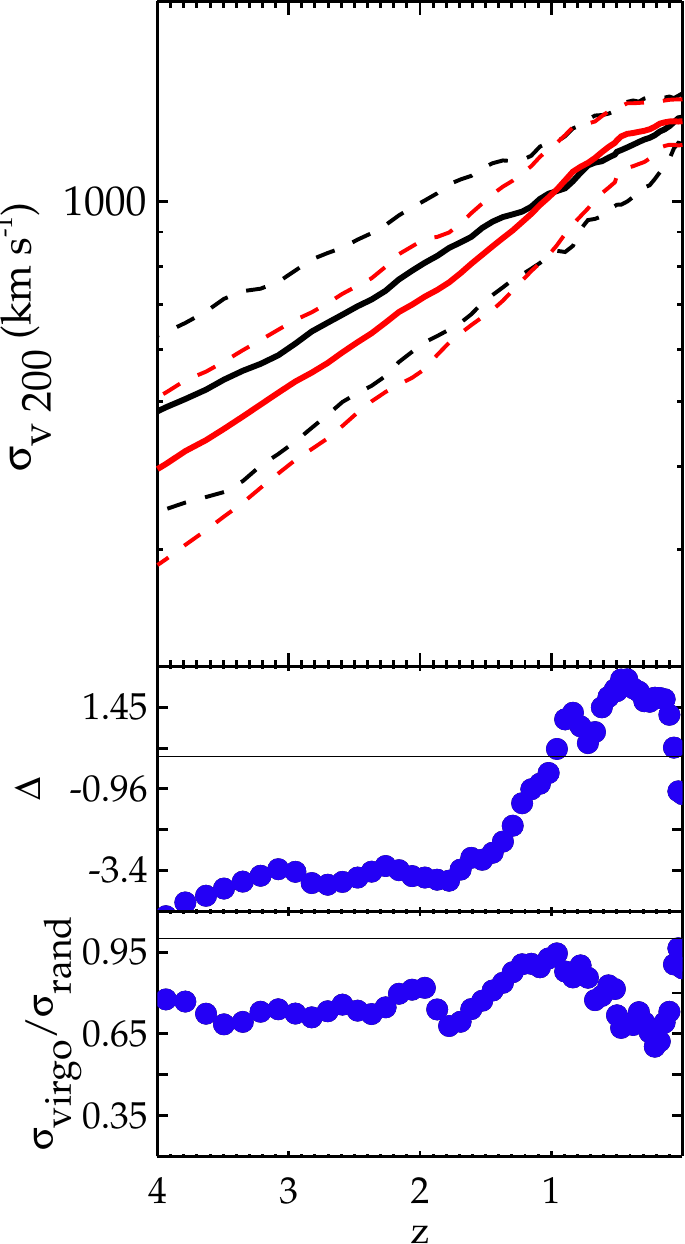}\hspace{0.5cm}
\includegraphics[width=0.177\textwidth]{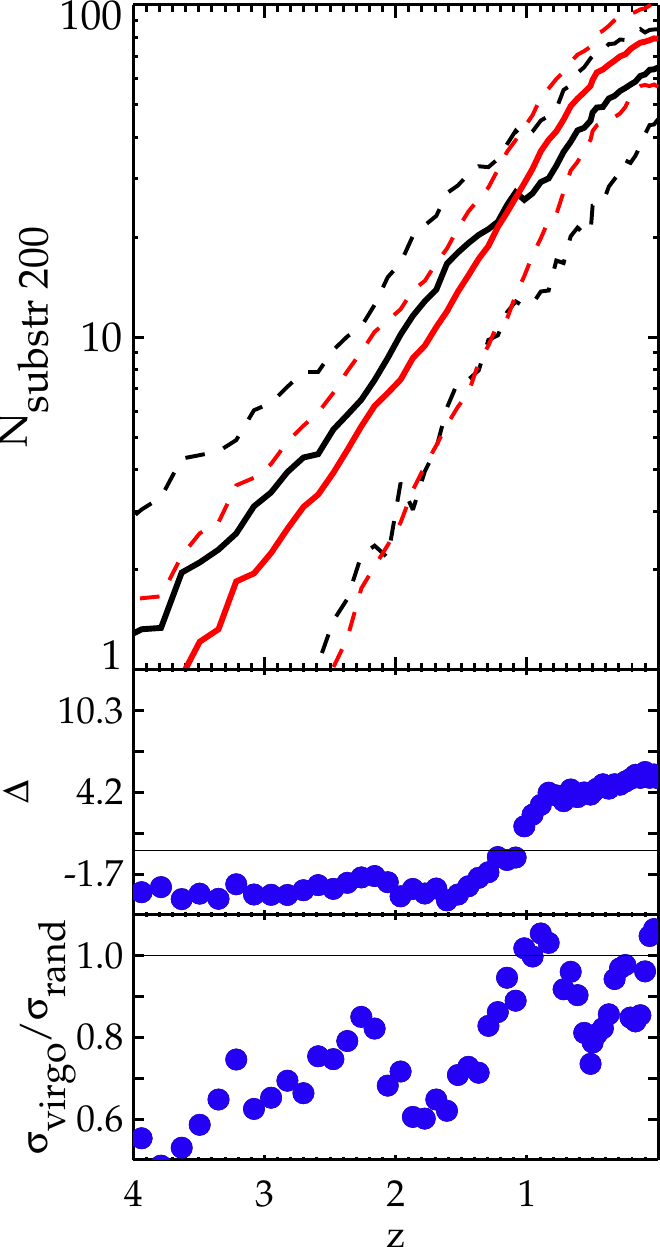} \\
\vspace{0.5cm}

\includegraphics[width=0.18\textwidth]{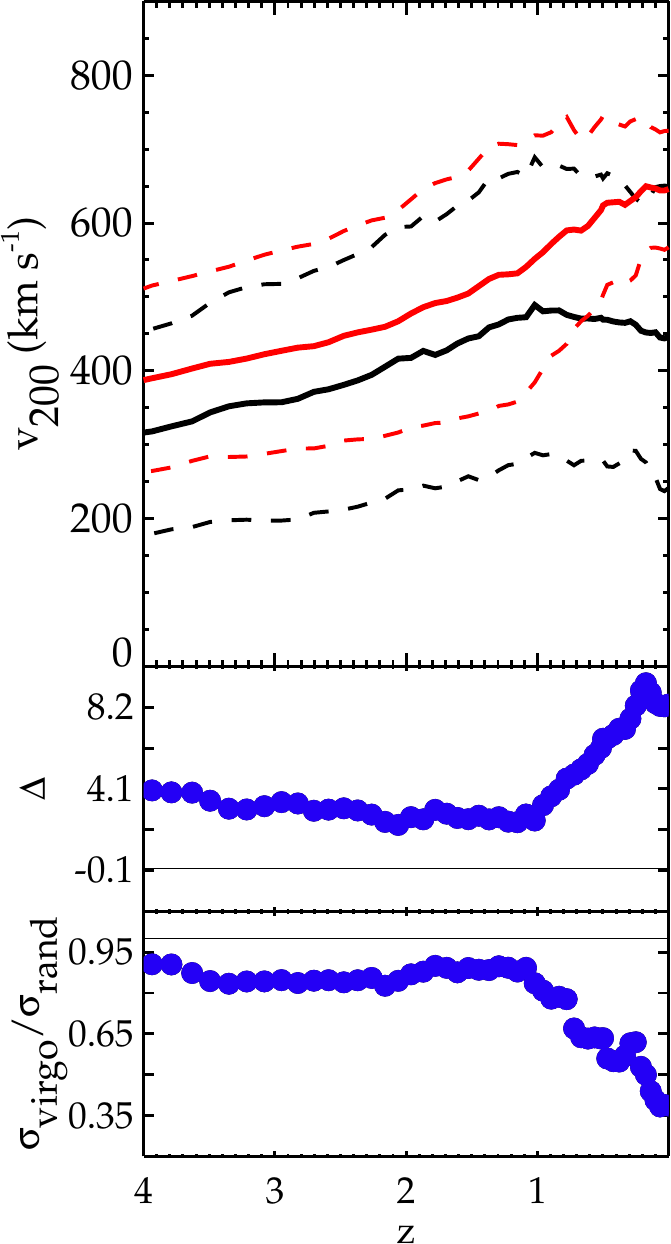}\hspace{0.5cm}
\includegraphics[width=0.18\textwidth]{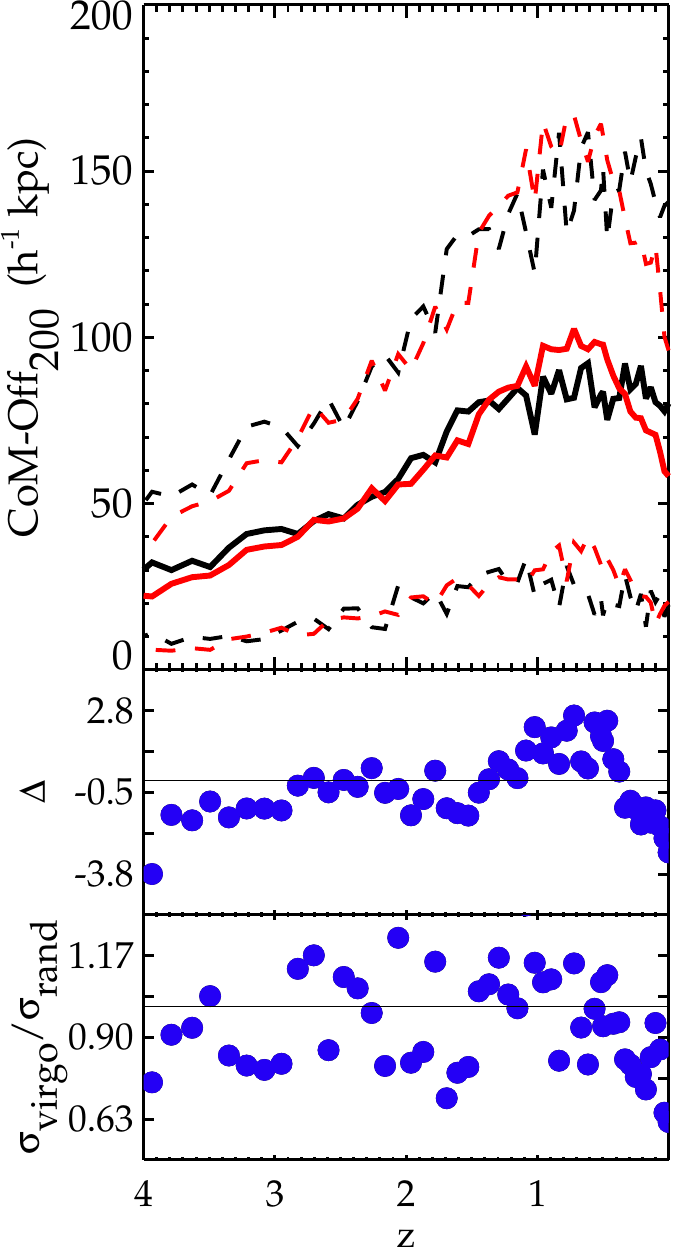}\hspace{0.5cm}
\includegraphics[width=0.18\textwidth]{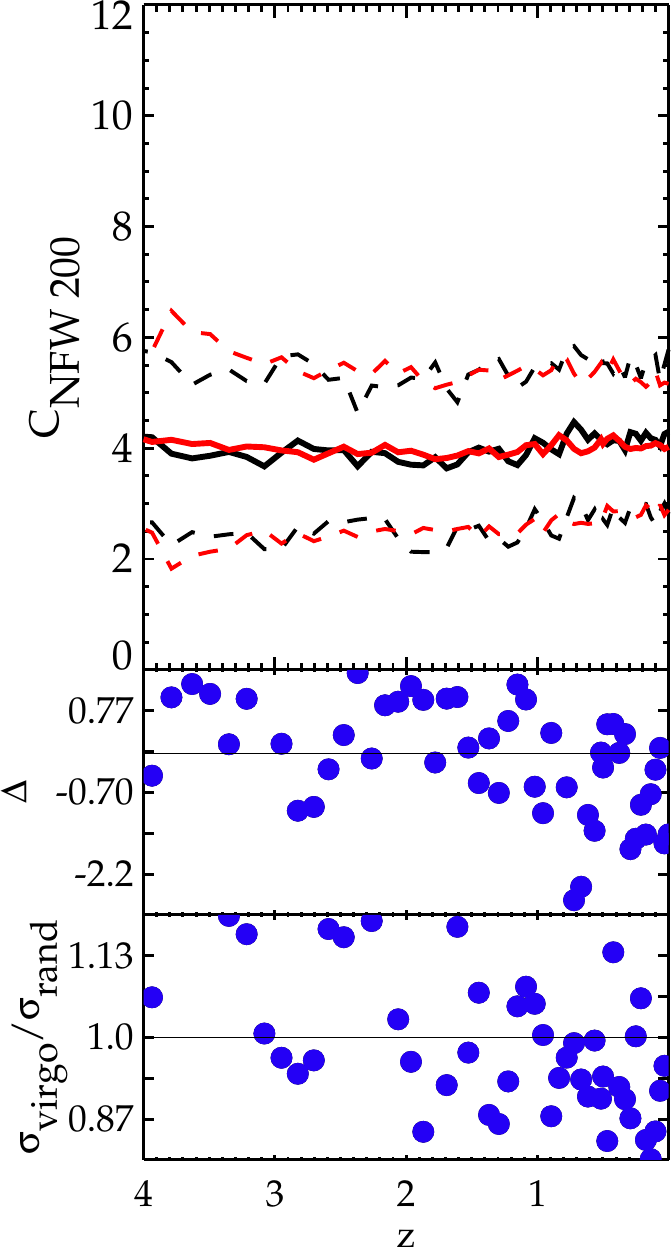}\hspace{0.5cm}
\includegraphics[width=0.18\textwidth]{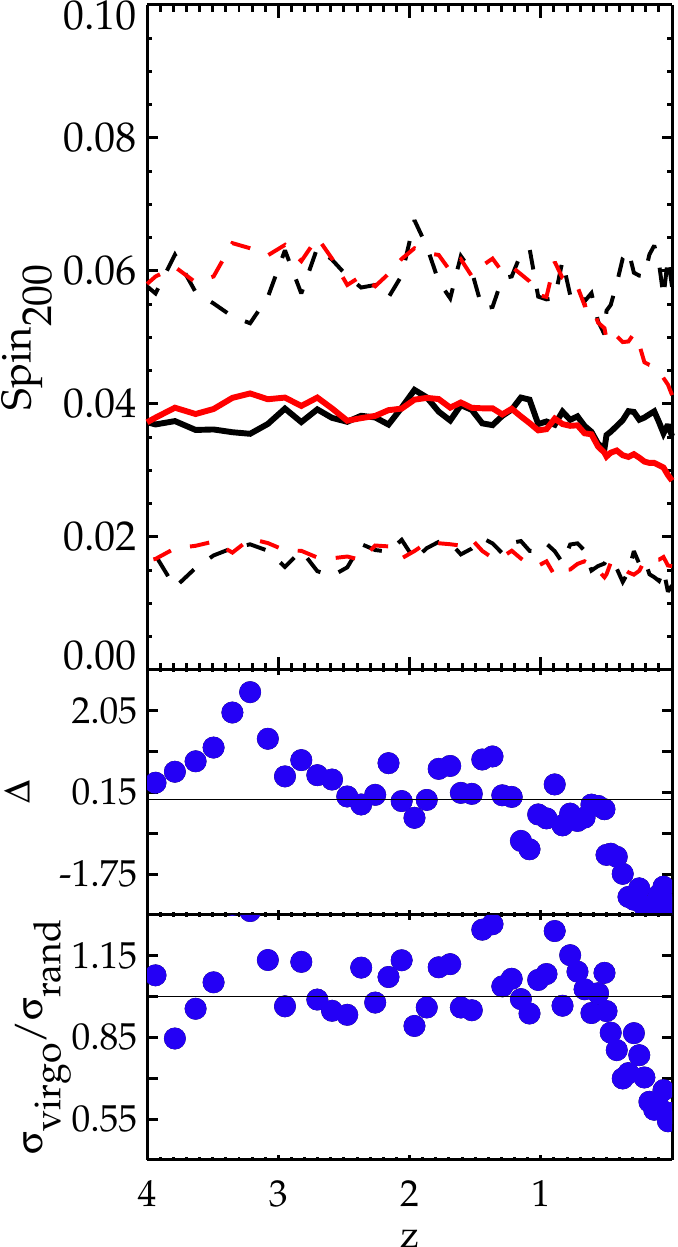}\hspace{0.5cm}

\caption{Top row: average property (solid line) and its standard deviation (dashed lines) as a function of the redshift for the 200 Virgo halos (red) and for the 77 random halos with M$_{0}^{\rm s}$ (black). Middle row: quantitative difference between the mean random and constrained property derived with the formula given in equation \ref{eq:1}. The solid black line denotes the value of insignificant differences (see the text for a detailed explanation). Bottom row: ratio of the standard deviation of the constrained property to that of the random property. The solid black line delimits the constrained zone (value inferior to 1) from the unconstrained area (value greater than 1). From left to right, top to bottom, properties are mass, radius, velocity dispersion, number of substructures, velocity, center of mass offset with respect to the spherical center, concentration and spin.}
\label{fig:nsubstr}
\end{figure*}

\begin{equation}
\Delta=\frac{\overline{\rm X_{\rm virgo}}-\overline{\rm X_{\rm rand}}}{\sqrt{\sigma_{\rm virgo}^2/n_{\rm virgo}+\sigma_{\rm rand}^2/n_{\rm rand}}}
\label{eq:1}
\end{equation}
where $n$ is the number of halos in the considered sample, $\sigma$ the standard deviation and $\overline{X}$ the mean. $\Delta$ is given in standard error units. Typically, definitions used in the following are : 1) if $\Delta$ in absolute values is larger than 3 (9) standard error units then it means that the two means differ significantly / at 99.9\% (extremely significantly / quasi at 100\%) ; 2) If the ratio of the standard deviations is smaller than 0.95 (0.55), the range of possible values for a given property of Virgo-like halos is divided by more than 1.05 (1.8) and the property is constrained (extremely constrained).\\

In the following, in addition to the random/constrained comparison at redshift zero visible on Figure \ref{fig:proba}, since the evolution might tell us more about the typicalness of the Virgo cluster, we further probe also the links between the properties at redshifts higher than zero. We thus examine one property at a time for both the Virgo sample and the mass-unbiased random sample. Tentatively, properties are gathered in two groups: 1) the internal properties that are a priori linked namely the mass, radius, number of substructures, velocity dispersion and to a lesser extent the concentration and 2) the external properties due a priori purely to the environment i.e. velocity, spin and the center of mass offset with respect to the spherical center.  
 
\begin{itemize}
\item Number of substructures:  The top row of the rightest top plot in Figure \ref{fig:nsubstr} shows that, before redshift 1, Virgo halos have slightly less substructures than random halos and after redshift 1 they have more substructures than random halos. Apart at redshift 1, Virgo halos tend to be atypical ($\Delta$$\neq$0). The fourth row of Figure \ref{fig:proba} confirms that Virgo halos have on average more substructures than random halos with M$_{200,0}^{\rm s}$ given their mass, radius and velocity dispersion. The enlarged panel establishes the absence of mass biases between the random and Virgo samples. There is a clear signal of about 1$\sigma$. It can be attributed to the fact that the Virgo cluster is a priori not fully virialized. Some internal structures are still in the process of merging \citep{2014A&A...570A..69B}. In addition, at every redshift larger than 1, the last row of the left plot in Figure \ref{fig:nsubstr} shows that the possible range of number of substructures for Virgo halos is about 10-20\% (or even 30-40\%  in some cases) smaller than that for random halos. \\

\item Mass and merging history: According to the top row of the leftest top plot in Figure \ref{fig:nsubstr}, the mean merging history of the Virgo halos is different from that of the selected random halos as already noticed in \citet{2016MNRAS.460.2015S} but with a much smaller sample. They differ significantly ($\lvert\Delta\rvert~>$~3) at redshift earlier than 1. Again redshift 1 seems to be a turning point for the Virgo halos. At this redshift, their mass accretion rate decreases drastically. At later redshifts, the random and constrained mean merging history differ moderately ($\lvert\Delta\rvert~\sim$~1). It means that for their mass, at all redshifts, the Virgo halos do not have a typical merging history. They had a more active history in the past than within the last 7 Gigayears. This observation is in agreement with the halo-bias at fixed mass: a halo that grows in an underdensity accreted a large amount of matter at high redshifts. While somewhat boosted by another cluster in the past, the latter has to be not too close for the accretion of the cluster of interest nowadays to be very weak but not stalled \citep{2009MNRAS.398.1742H,2017MNRAS.469..594B,2018MNRAS.476.4877M}. The Centaurus cluster is indeed in the vicinity while still about 15~\hMpc\ away from Virgo and the local Universe seems to be a general underdensity \citep{2013ApJ...775...62K}. The link between the atypicality of the Virgo cluster merging history and its environment is established.  The different past histories of the clusters are tightly entangled to their environment. In addition, except at redshift about 1, the Virgo halos have a smaller range of possible merging histories than random halos (ratio of the standard deviations smaller than 1). These assessments underline that getting clusters with a merging history matching that of the Virgo halos requires several criteria.\\
 
\item Radius (Figure \ref{fig:nsubstr}, second top panel): The radius presents the exact same behavior as the merging history, a completely expected behavior because of its strong correlation with the mass. \\

\item Velocity dispersion  (Figure \ref{fig:nsubstr}, third top panel): The velocity dispersion also strongly correlated to the mass is no exception. Redshift 1 appears again as a turning point.\\

\item Concentration (Figure \ref{fig:nsubstr}, third bottom panel): It is a commendable exception as it is stable across late cosmic time for both constrained and random halos without major differences between the two. In that respect, Virgo halos are typical at all redshifts and not only at z=0.\\
\end{itemize}

\begin{itemize}
\item Center of mass offset: The top row of the second bottom plot in Figure \ref{fig:nsubstr} shows the evolution of the offset of the center of mass of the halos with respect to their spherical center. For both random and constrained halos, this offset increases until a redshift approximately equal to 1. At smaller redshifts than 1, the offset decreases more or less rapidly and faster for the Virgo halos than for the random halos. The quiet merging history of the Virgo cluster in the last few gigayears might be the reason for such an observation. The offset tends to be slightly larger for Virgo halos than for random halos at redshifts close to 1. At redshift zero, constrained mean offset values are smaller than that for the random halos in a significant way (second row of the plot). The ratio of the standard deviation is centered on 1. Namely the offset of the center of mass is overall not constrained according to the definition chosen at the beginning of the section where we split between constrained and atypical attributes.  However, interestingly Figure \ref{fig:proba} shows a stronger correlation between the center of mass offset and the mass (or radius, velocity dispersion, number of substructures) for the Virgo halos than for the random halos. The indirect constraint on the merging history (constraint via the large scale environment) is probably the reason for such an observation. Basically, although there is still a residual cosmic variance in terms of mass, at a given mass the constrained history tends to favor lower offsets with respect to the center of mass. However, with increasing mass, this offset still rises irremediably because the most massive Virgo halos must accrete more mass although according to the same scheme as the smallest ones in the same amount of time.\\

\item Spin (Figure \ref{fig:nsubstr}, last bottom panel): The spin behaves exactly like the other parameters with the change of trend at redshift 1. Note that the random and constrained mean values of the spin are always only moderately different. The ratio of the standard deviations of the constrained and random spin values is smaller than 1 by up to 45\% for redshift smaller than 1. Figure \ref{fig:proba} confirms that at redshift zero the spin of a Virgo halo has a smaller range of possible values than an average random halo. Since spin and local environment are linked, it confirms that Virgo halos are in a well constrained environment at least within the last 7 Gigayears with no major merger.\\

\item Velocity: The behavior of the mean velocity is shown in Figure \ref{fig:nsubstr}, first bottom panel. As expected, masses and velocities have no strong link. Hence, while the random halos have an average velocity of 463$\pm$207~\kms, the Virgo halos have an average velocity higher by almost 200~\kms\ (646$\pm$79~\kms\ barely within 1$\sigma$ of the random distribution)\footnote{Note that the offset that may be induced by the different boxsizes is smaller than the standard deviation of the constrained velocities \citep{2003MNRAS.339..271S}} at redshift zero (cf. Figure \ref{fig:proba}). Typically, at all redshifts, the Virgo halos have a larger velocity than the random halos (top row) with a high level of significance (second row) and different standard deviations (third row). The latter differ significantly between redshifts 0 and 1 (last row). The ratio of the constrained and random standard deviations is always smaller than 1. As a side note, the average velocity vector direction of the Virgo halo is (-529$\pm$78 ; 268$\pm$66 ; -236$\pm$78)~\kms. As expected, the vector points in the direction of the Great Attractor and beyond, where the Shapley supercluster stands.
\end{itemize}

This analysis shows us that 1) the Virgo cluster is a priori not a completely typical cluster especially regarding its velocity, or in other words its large scale environment is not typical thus it gives birth to an atypical cluster, 2) some characteristics of the Virgo cluster are statistically constrained not only at redshift zero but also at earlier redshifts. The next section aims at summarizing all these findings.

\subsection{Discriminative properties of the Virgo cluster}

\begin{table}
\begin{center} 
\begin{tabular}{l@{ }@{ }@{ }@{ }@{ }@{ }@{ }@{ }@{ }@{ }@{ }@{ }c@{ }@{ }@{ }@{ }@{ }@{ }@{ }@{ }@{ }@{ }@{ }@{ }c@{ }@{ }@{ }@{ }@{ }@{ }@{ }@{ }@{ }@{ }@{ }@{ }c@{ }@{ }@{ }@{ }@{ }@{ }@{ }@{ }@{ }@{ }@{ }@{ }c}
\hline
\hline
& (1) & & (2) &  \\
\hline
\hline
R$_{200}$            & -0.44$\pm$0.21 &  & 0.91$\pm$0.03  &-- \\
$\sigma_v$          &   -0.79$\pm$0.28 & & 0.87$\pm$0.04 &--  \\
N$_{\rm substr}$  & 5.36$\pm$0.14 &-- & 1.2$\pm$0.04 &   \\
v                           & 8.10$\pm$0.53 &--  & 0.38$\pm$0.01 &\fbox{ }  \\
CoM-off                &  -2.69$\pm$0.32  & &  0.62$\pm$0.01 &-- \\
C$_{\rm NFW}$   &   -1.40$\pm$0.40 & & 0.81$\pm$0.03 &-- \\
spin                     &   -2.30$\pm$0.41 &  & 0.62$\pm$0.05 &--  \\
\hline
\hline
\end{tabular}
\end{center}
\vspace{-0.25cm}
\caption{(1) Difference between the mean property values of the 200 Virgo halos and of the selected random halos / (2) Ratio of the standard deviations of their properties. 1,000 different draws of halos among the cluster-size random sample have been made to estimate the variance of $\Delta$ and the ratio on the random sample drawn to math the mass distributions. A long dash (frame) highlights an (a highly) atypical / constrained property value (according to equation \ref{eq:1}, absolute values above 3 and 9 in standard error units / values below 0.95 and 0.55 respectively) for the Virgo halos with respect to selected random halos.}
\label{Tbl:3}
\end{table}

Table \ref{Tbl:3} recapitulates the properties studied in the previous section and highlight those atypical and/or constrained, i.e. that differ (significantly) between Virgo and selected random halos at redshift zero. As previously stated, the atypical and constrained adjectives can qualify a property if : \#1 the constrained and random mean values of the property differ (significantly). Namely, equation \ref{eq:1} gives a non zero and large absolute value ; \#2 the constrained and random property distribution have (significantly) different standard deviations with a ratio constrained to random smaller than 1. $\Delta$ and ratio values are followed either by a long dash (\#1 $\lvert\Delta\rvert~>~$3 ; \#2 ratio$<$0.95) or a frame ((\#1 $\lvert\Delta\rvert~>~$9 ; \#2 ratio$<$0.55) in the table to highlight the different or extremely different values. The standard deviations of these values are obtained by drawing 1,000 times halos among the cluster-size halos to get samples that match the mass distribution of Virgo halos at z=0. We give hereafter examples of properties fulfilling condition \#1, condition \#2 or both conditions: 

\begin{itemize}
\item if the random halos are selected according to their mass (column (1) of Table \ref{Tbl:3}), then Virgo halos are atypical in terms of their number of substructures and velocity in the sense that their mean values fulfill condition \#1.   
\item According to condition \#2, although Virgo halos have overall typical center of mass offset, concentration and spin values, their range of possible values is smaller than that of the random halos selected upon their masses as shown in column (2) of Table \ref{Tbl:3}. 
\item The line (5) of table \ref{Tbl:3} shows that the velocity fulfills the two conditions. This is not unexpected since the simulations all resemble the local Universe. The reproduced local large scale structure then induces the same motion for all the Virgo halos. However, the low probability for random halos to have the same motion as Virgo halos indicates the low probability to have a local environment inducing this velocity.
\end{itemize}

According to this table the most atypical property of the Virgo halos at z=0 is their velocity. Their number of substructures is also not quite typical but to a lesser extent. If given their masses, Virgo halos have globally more substructures than random halos it is not as significant as the atypicality of their velocity. On the contrary, the concentration is the less atypical value and the number of substructures is the less constrained one, although a more detailed study is required here to quantify the masses of these substructures.\\

\begin{figure}
\hspace{0.5cm}\includegraphics[width=0.35\textwidth]{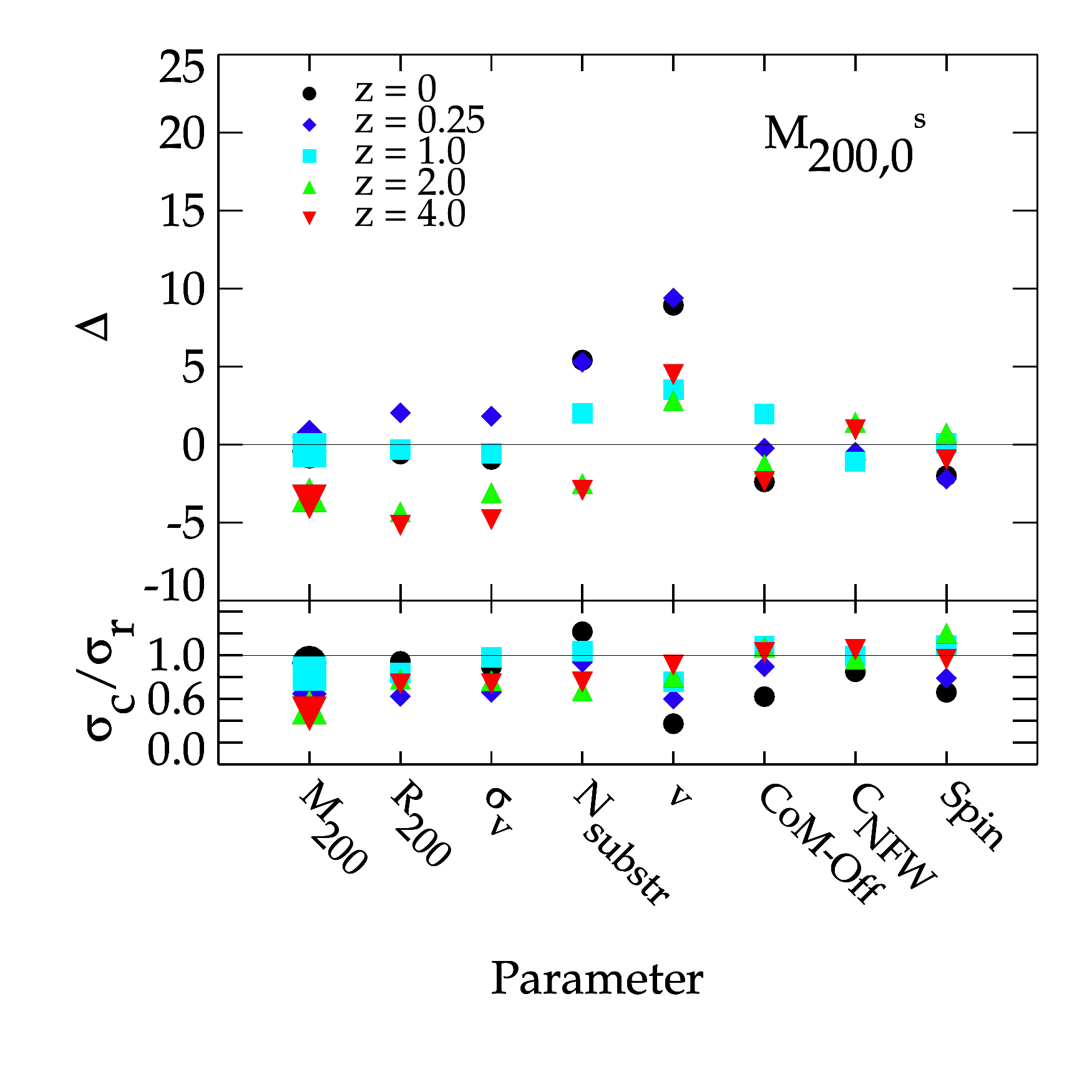}\\
\vspace{-0.6cm}
\caption{Quantitative measurement of the difference between the mean property values of the 200 Virgo halos and of the selected random halos (top row, equation \ref{eq:1}) as well as the ratio of their standard deviations (bottom row). The mass distribution of the random halos is matched (thus plotted as the largest symbol) to that of the Virgo halos. Measurements are given at different redshifts (colored symbols) in all the panels.}
\label{fig:summary1}
\end{figure}

This table is valid at redshift zero. It would be fastidious to present it at all redshifts. Instead, Figure \ref{fig:summary1} gives a visual overall impression of the status (atypical or/and constrained) of the properties at different redshift. $\Delta$ and the ratio of the constrained and random standard deviations are derived for each property under study at different redshifts (one color per redshift).\\

Links between the different properties are visible. For instance,as expected the radius is quite typical. More importantly redshift 1 appears clearly as the redshift of changes: the light blue filled squares stand out between the symbols used for the earliest redshifts and those used for the latest redshifts.  For instance, R$_{200}$, $\sigma_v$ and N$_{\rm substr}$ are atypical at z$>$0 but at z=1: from large values at 0$<$z$<$1, $\Delta$ becomes quasi-null at z$=$1 to increase again in absolute value. To conclude, most of Virgos' properties appear to be atypical and/or constrained at various levels at all redshifts but 1. \\

So far, random halos have been selected only on their mass and an additional criterion was added to their mass distribution. It is interesting to take a step back to see how many random halos have property values that all fall within 3$\sigma$ of Virgos' values at redshift zero without the additional criterion, i.e. considering the 400$+$ random halo sample. We first check how many of the Virgo halos are in their own 3$\sigma$ scatter simultaneously for the 8 properties under study in this paper. 97\% of the Virgo halos comply with this request. By comparison, only 30\% of the random halos are left. Virgo is thus an unlikely cluster of galaxies nowadays and by extension at earlier redshifts: until at least redshift 1, the redshift of the changes. \\

\begin{figure}
\vspace{-0.6cm}
\centering
\hspace{-0.8cm}\includegraphics[width=0.2\textwidth]{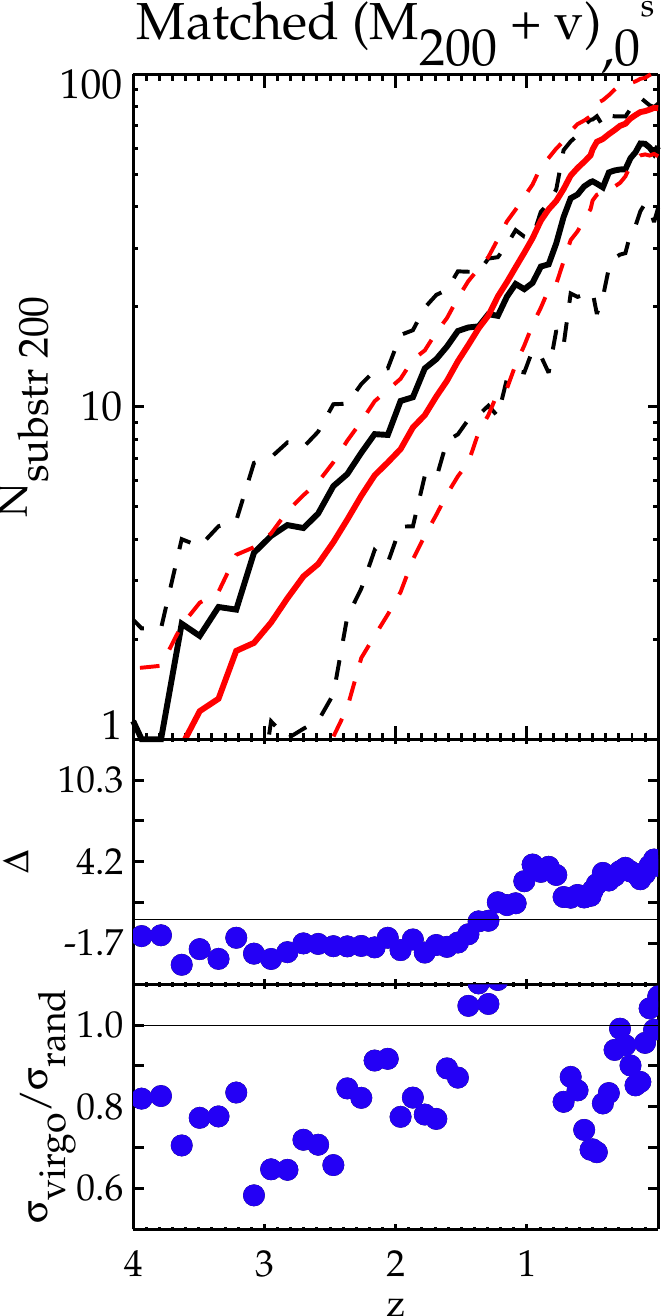}
\vspace{-0cm}
\caption{Same as Figure \ref{fig:nsubstr} but random halos are selected so as to match both their mass and velocity distributions to those of the Virgo halos.}
\label{fig:vllI}
\end{figure}

To understand which property is responsible for this huge drop in the number of halos besides the constraint on their mass (they need to be cluster-size halos), it is then interesting to add an additional selection criterion. This additional criterion consists either in matching successively a given parameter distribution of the selected random halos to that of the Virgo halos or in simply restricting the range parameter values to [$\overline{\rm X_{0,\rm virgo}}$-2~$\times~\sigma_{\rm Xo,virgo}$ , $\overline{\rm X_{0,\rm virgo}}$+2~$\times~\sigma_{\rm Xo,virgo}$]  where $\sigma_{\rm Xo,virgo}$ is the standard deviation of the Virgo-like X property and $\overline{X_{0,\rm virgo}}$ is the mean X property of the Virgo-like halos at z=0. The appendix gathers the results for the different additional selection criteria. Selecting the cluster-size random halos on the additional velocity distribution criterion is the major cause of decrease in the number of random halos left (16\%, 49\% restricting only the range of velocity without reproducing the distribution). This observation then tends to imply that our local environment is quite atypical.
Moreover, Figure \ref{fig:vllI} shows that two additional criteria to match the cluster-size random halos' velocity and mass distributions to that of the Virgo halos is not sufficient to get the proper number of substructures. \\

Furthermore, adding the third criterion consisting in match the distribution of the substructures numbers of the random halos to that of the constrained halos random halos does not give solely random halos that have a quiet merging history within the last 7 gigayears (i.e. they had a major merger within the last 7 gigayears, their masses grew rapidly). With the more relax range restriction rather than distribution matching criterion on mass, velocity and number of substructures, only 18\%(0.5\%) of these random halos have merging histories within 3(2)$\sigma$ of that of the Virgo halos. None are within 1$\sigma$. This implies that merging histories are not defined solely by the properties under study in this paper (although the velocity linked to the environment is considered). Reversely, the merging history type constrains some parameters. Studying the impact of the environment on merging histories to find the type of environments leading to merging histories similar to Virgo halos will shade some light on the environment we are living in. The essential peculiarities of the latter for having a Virgo cluster with such a merging history can be highlighted. Such a study that requires a huge halo sample for statistical purposes is underway. It will also focus on the redshift of change, z=1. Actually, several turning points are known to occur during the formation of the dark matter halos for different reasons, like for instance when their concentration reaches a given value at z$>$4 \citep{2003MNRAS.339...12Z}. This paper highlights yet another turning point that occurs at about half the Universe age indicating that there are most probably, in a first approximation, two types of environment giving two types of halo merging histories intersecting at about z=1. The mean merging history of the random halos gives the average between two types of merging histories: one like the Virgo halos with major accretion in the past and a recent slow down, another with minor accretion in the past and a recent increase. However, the current number of random halos is insufficient to confirm definitively this hypothesis. Still, note that this hypothesis and the observations are in full agreement with the halo-bias at fixed mass described in 4.1.

\section{Conclusion}

Galaxy clusters are powerful cosmological probes. Combined with their numerical complement that emerged within the past few years, they permit studying the formation and evolution of clusters and testing theoretical models. As such Virgo constitutes a formidable source of information via detailed observations facilitated by its proximity with us. However, accessing all the properties and the past history of the cluster from nowadays observations to select the numerical lookalike valid for a one-to-one detailed comparison is not completely trivial. A wide variety of existing clusters, most probably a result of an environmental diversity, complicates the detailed comparisons on a one-to-one basis.\\

At this stage, accurate lookalikes of the Virgo cluster come in handy to determine the properties of the Virgo cluster, its likeliness and thus its environmental likelihood as well as to supply the numerical complement. We obtain 200 of such Virgo-like halos by applying the zoom-in technique to 200 constrained initial conditions. By definition of a constrained simulation, the Virgo halos formed in a reproduction of our local environment. These simulations have been proven in the past (but only with a very limited sample size) to give accurate reproductions of the Virgo cluster in general (i.e. mass, position and merging history). Besides conducting a thorough statistical study of these 200 halos, we compare them to random halos extracted from a set of 3 random simulations with the same (resolution and cosmological framework) features as those used for the Virgo halos. \\

Random halos are selected so that their mass distribution reproduces that of the Virgo halos. Results and conclusion are unaffected by the set of selected random halos. Two appellations are used to characterize a given property: a property can be considered 1) atypical or/and 2) constrained. It is atypical if its average value obtained for the Virgo halos differ (significantly) from that obtained for the random halos. It is constrained if its standard deviation is (significantly) smaller for Virgo halos than for random halos. Studies are conducted at different redshifts. \\

Conclusions are as follows for a set of random halos sharing the same mass distribution as that of the Virgo halos:
\begin{itemize}
\item As expected by definition of a constrained simulation valid down to the cluster scale, most of the properties of the Virgo halos are constrained with respect to those of the random halos at all redshifts. The only exception is the number of substructures for redshift smaller than 1. A more detailed analysis of these substructures (masses, positions) will follow.
\item Until about 7.8 Gigayears ago (z=1),  the trends of most properties of the Virgo halos with respect to those of the random halos were reversed with respect to today: property values that were larger (smaller) for Virgo halos than for random halos became smaller (larger) at z=1. Hence, Virgo halos have on average a larger number of substructures, a quieter merging history nowadays than random halos, sharing the same mass distribution as them at z=0, while it was the opposite at redshifts larger than 1. The environment is probably the cause for such an observation. This reinforces the necessity to simulate clusters in the proper environment to push further the comparisons with observed ones.
\item Required but not sufficient criteria to select random halos that match Virgo halos are the mass (or alternatively the radius) and the velocity at z=0. These two criteria ensure that radius (mass), velocity dispersion, concentration and spin are shared between random and Virgo halos at redshift zero on average. However, the offset of their center of mass with respect to the spherical center is not constrained and their numbers of substructures are about 1$\sigma$ away. Obviously the random merging histories are not all quiet. The velocity is a necessary but not sufficient condition to get the proper environment even if it is correlated on large scales.
\item A third requirement to select a perfect candidate for the Virgo cluster will then be for instance the number of substructures. However, even if the random halos have values within 3$\sigma$ of the average Virgo values, the merging history of random halos still differs from that of Virgo halos on average. These selection criteria are efficient only at z=0. The merging history itself can then be considered as a criterion.
\item Nevertheless, getting a random halo that matches the Virgo cluster is a daunting task. Only 18\%(0.5\%) of the random halos have their 8 properties (mass, radius, velocity dispersion, number of substructures, concentration, offset with respect to the center of mass, velocity and spin) within 3(2)$\sigma$ of the average Virgo values at z=0 and their merging history that matches within 3(2)$\sigma$ that of Virgo halos up to redshift 4. None are within 1$\sigma$. Since the merging history forged by the environment is the additional requirement, it becomes clear that simulated clusters are accurate counterparts of the observed ones only when they are simulated in the proper large scale environment. 
\end{itemize}

Consequently, 1) this set of 200 Virgo halos highlights the complexity in getting a numerical cluster that matches the observed Virgo cluster of galaxies. Such a pairing is however essential to push studies further in details. This paper gives the values of the properties to be matched especially if they affect the galaxy population of the cluster like the merging history does \citep[e.g.][]{2015ApJ...807...88G}. Only then detailed comparisons between simulated and observed galaxy populations are legitimate to test galaxy formation and evolution models and to calibrate them~; 2) this set of 200 Virgo halos also opens great perspectives to lead comparisons with observations as they simplify grandly the challenge of selecting the proper simulated cluster. In particular, it will permit studying the substructures of the Virgo cluster. The impact of the environment on cluster properties can be studied in more details. The reason for the quiet merging history of the Virgo cluster within the last seven gigayears already previously highlighted but with a much smaller statistical sample of Virgo halos certainly deserves attention. An ongoing study with an extremely large statistical sample of halos suggests a potential link between the type (quiet/active) of merging histories and the number of neighbors (depending on their distance and mass) at z=0 and may also explain the existence of a redshift of change. Finally, a few representative Virgo halos will be selected from this set to run zoom-in hydrodynamical simulations of the Virgo cluster that will permit testing precisely the galaxy formation and evolution models giving birth to galaxy cluster populations. In addition, the implications of having a massive neighbor for the local Group, in particular our Galaxy, might be highlighted.\\


\section*{Acknowledgements}

The authors gratefully acknowledge the Gauss Centre for Supercomputing e.V. (www.gauss-centre.eu) for providing
computing time on the GCS Supercomputers SuperMUC at LRZ Munich. JS thanks her CLUES collaborators in particular Stefan Gottl\"ober for useful discussions.
JS acknowledges support from the ``l'Or\'eal-UNESCO Pour les femmes et la Science'' and the ``Centre National d'\'etudes spatiales (CNES)'' postdoctoral fellowship programs.

\begin{table*} 
\begin{center} 
\begin{tabular}{lrrrrrr@{ }@{ }@{ }@{ }rr}
\hline
   \hline
Additional set parameter & mass & radius &  \multicolumn{1}{c}{number of}  & \multicolumn{1}{c}{velocity}&\multicolumn{1}{c}{velocity} &  \multicolumn{1}{c}{center of mass} &  \multicolumn{1}{c}{concen-} & spin  \\
distribution at z=0     &             &          & substructures & dispersion & &  \multicolumn{1}{c}{offset wrt to the}   & \multicolumn{1}{c}{tration} \\
 ( X$_0^{\rm s}$ )           &		&		&		&		& & \multicolumn{1}{c}{spherical center} & & \\
 \hline
\hline
Random &77& 82 & 75 & 101 & 66  & 259 & 374 & 359 \\ 
\hline
\hline
\end{tabular}
\end{center}
\vspace{-0.25cm}
\caption{Number of cluster-size random halos available for comparisons with the 200 constrained halos for each parameter distribution that is chosen to be set at redshift 0. Namely, we select only random halos with a parameter, X$_{0,\rm rand}$, such that it belongs to [$\overline{\rm X_{0,\rm virgo}}$-2~$\times~\sigma_{\rm Xo,virgo}$ , $\overline{\rm X_{0,\rm virgo}}$+2~$\times~\sigma_{\rm Xo,virgo}$]  where $\sigma_{\rm Xo,virgo}$ is the standard deviation of the X$_{0,\rm virgo}$  and $\overline{\rm X_{0,\rm virgo}}$ is their mean at z=0. In addition, random halos in the tails of the distribution are only partially selected so that random and constrained distributions match each other.}
\label{ATbl:1}
\end{table*}

\section*{Appendix}
\renewcommand{\thesubsection}{\Alph{subsection}}
\subsection{Other additional parameter distribution selection criterion}

Rather than matching the mass distribution of the cluster-size random halos to that of the Virgo halos, it is possible to match other parameter distributions, namely to use another additional selection criterion besides the mass range. Each one of the subsamples obtained from the cluster-size random halo sample must reproduce one of the parameter distribution at redshift zero (X$_0$,  the only one that can be compared with the observational value measured for the Virgo cluster when possible) of the Virgo lookalike sample. The set parameter distribution of the random halo subsample is denoted X$_{0}^{\rm s}$ (always set at redshift zero).\\

Such an additional parameter selection criterion can give an idea of the commonness of the Virgo cluster with respect to that parameter within a group of its peers (cluster-size halo). Table \ref{ATbl:1} gives the number of random halos left after applying successively the different X$^{\rm s}_0$ additional selection criterion. One can notice that clearly imposing the same velocity distribution to the random halos as that of the Virgo halos reduces drastically the number of cluster-size random halos available for comparisons (66 random halos out of 423 are left, i.e. 16\%). On the contrary, the NFW concentration has no serious effect (374 random halos left, i.e. 88\%) while the center of mass offset with respect to the spherical center has only a mitigated impact (259 random halos left, i.e. 61\%) on the number of random halos left. \\
Note that, although in any case random halos are selected to be within the same mass range as the Virgo halos or in other words to have a cluster size, some subsamples are biased toward the low mass (and all the other correlated parameters) end.\\
 
\begin{figure}
\vspace{-0.6cm}
\centering
\hspace{-0.8cm}\includegraphics[width=0.2\textwidth]{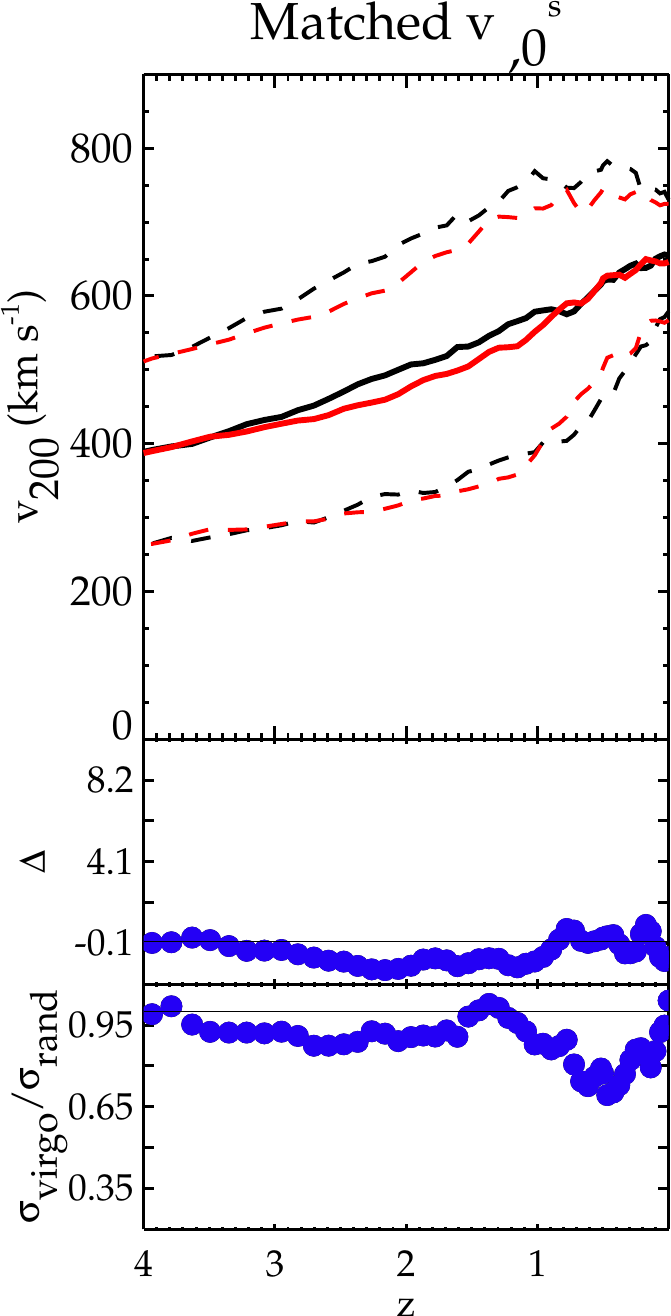}
\vspace{-0cm}
\caption{Same as Figure \ref{fig:nsubstr} but cluster-size random halos are selected so as to match their velocity distributions to those of the Virgo halos.}
\label{Afig:vlI}
\end{figure}

Figure \ref{Afig:vlI} shows the interesting case of the cluster-size random halos selected according to their velocity value. At all redshifts the random and constrained mean velocities present no significant differences. Interestingly, the ratio of the constrained and random standard deviations is smaller than 1 for redshift higher than about 0.5 up to redshift 1.  Note that once the velocity distribution is set to be the same at redshift zero for both the Virgo and cluster-size random halos, they stay similar at higher redshifts. This highlights the large scale correlation of the velocity field. It explains why the linear Reverse Zel'dovich Approximation used to put back constraints at the position of their precursors works well.\\

Tables \ref{ATbl:3} and  \ref{ATbl:4}  recapitulate the properties studied in the paper and highlight those atypical and/or constrained at redshift zero depending on the additional selection criterion (see the main core of the paper for details). \\

\begin{figure*}
\hspace{-0.3cm}\includegraphics[width=0.35\textwidth]{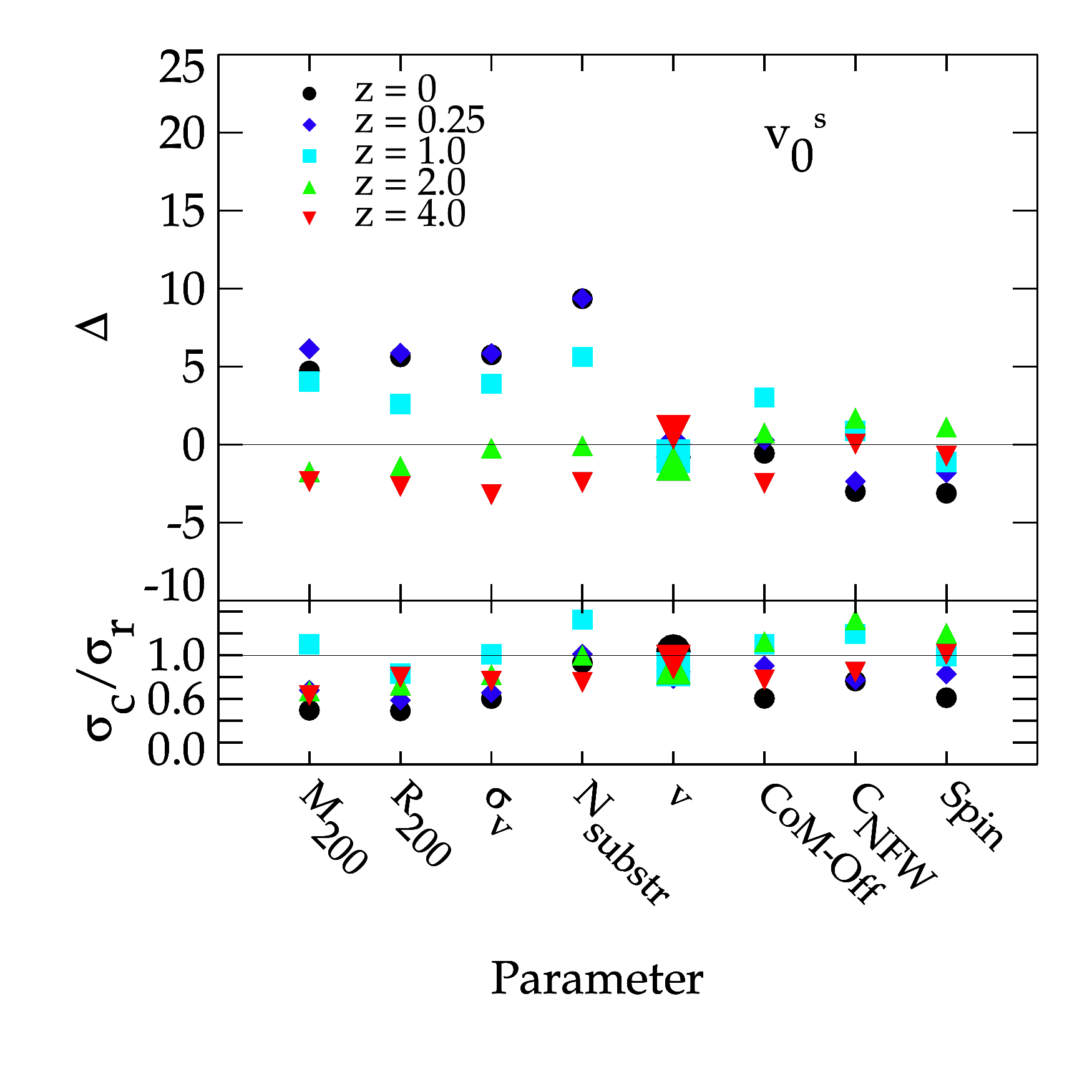}
\hspace{-0.3cm}\includegraphics[width=0.35\textwidth]{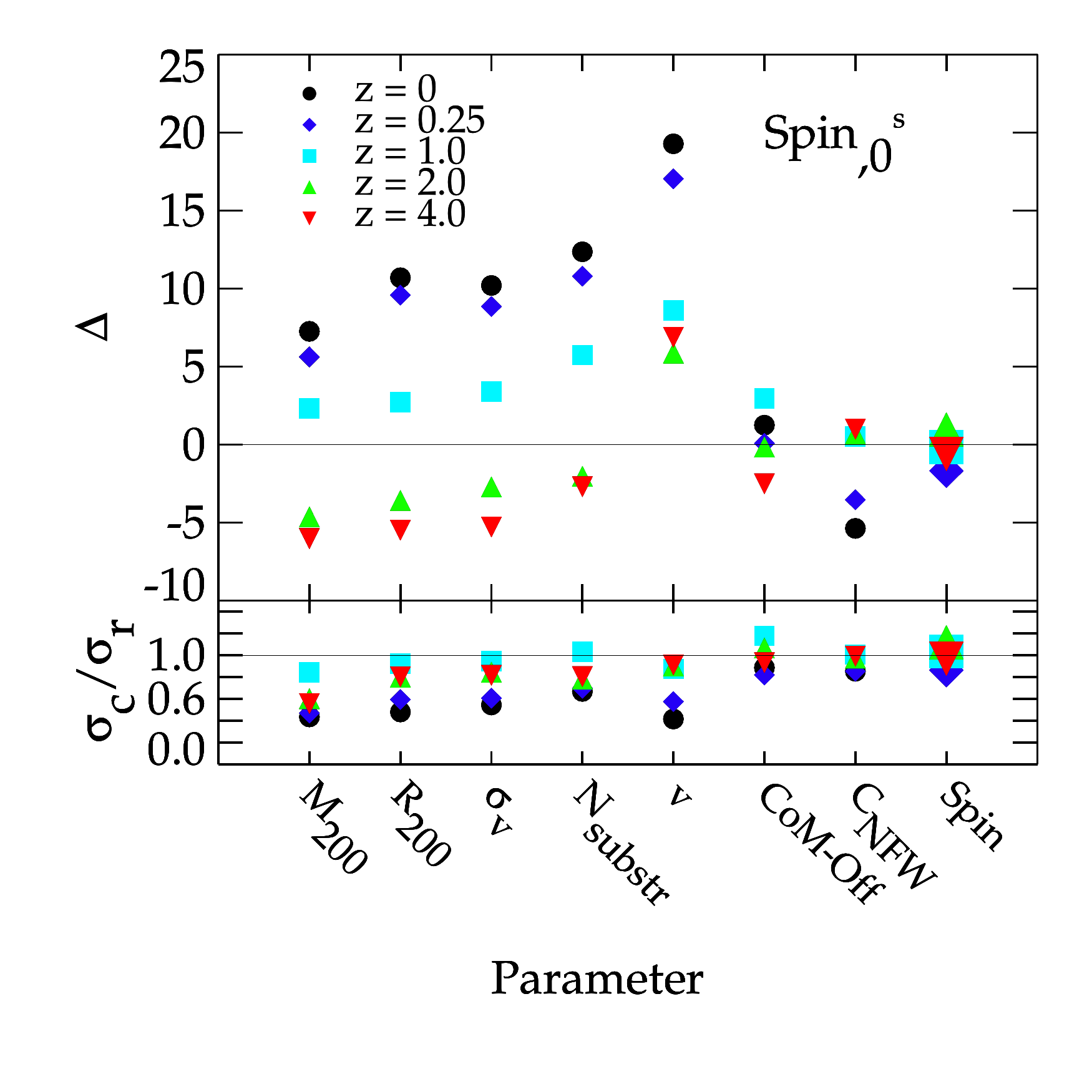} \\
\vspace{-0.6cm}
\caption{Each panel gives the quantitative measurement of the difference between the mean property values of the 200 Virgo halos and of the cluster-size selected random halos (top row, equation \ref{eq:1}) as well as the ratio of their standard deviations (bottom row). Cluster-size  random halos are selected according to the set parameter (one by panel)  at redshift 0 (name in the corner of each panel and largest symbol). Measurements are given at different redshifts (colored symbols) in all the panels.}
\label{Afig:summary1}
\end{figure*}

Finally, Figure \ref{Afig:summary1} corresponds to one X$_0^{\rm s}$ case identified in the top right corner of the panel and by the largest symbol.  The redshift of change stands up.

\begin{table}
\begin{center} 
\begin{tabular}{l@{ }@{ }@{ }c@{ }@{ }@{ }c@{ }@{ }@{ }c@{ }@{ }@{ }c@{ }@{ }@{ }c@{ }@{ }@{ }c@{ }@{ }@{ }c@{ }@{ }@{ }c@{ }@{ }@{ }}
\hline
\hline
&\multicolumn{8}{c}{X$_0^{\rm s}$: additional set parameter distribution at redshift 0}\\
& (1) & (2) & (3) & (4) & (5) & (6) & (7) & (8) \\
Param.  & M$_{200}$ & R$_{200}$ & $\sigma_v$ & N$_{\rm substr}$ & v & CoM-off & C$_{\rm NFW}$ & spin\\
\hline
\hline
M$_{200}$            & /  &     &       &          & --            &  --           &--          &-- \\
R$_{200}$            &    &  /  &       &           & --           & --             & \fbox{ } & \fbox{ }  \\
$\sigma_v$          &     &     &    /  &          & --           &  --            & \fbox{  } & \fbox{   } \\
N$_{\rm substr}$  & -- & -- &    --  &  /       & \fbox{   } &  \fbox{   } & \fbox{  } & \fbox{     } \\
v                           &--  &  --&   \fbox{   }& -- &  /          &   \fbox{      } & \fbox{ } & \fbox{  }  \\
CoM-off                &     &    & --   & --       &                &   /             &             &                    \\
C$_{\rm NFW}$   &     &     &      &          &                &   --           &  /          & --    \\
spin                     &      &    &   --   &      &                 &--              & --        &  /    \\
\hline
\hline
\end{tabular}
\end{center}
\vspace{-0.25cm}
\caption{Difference between the mean parameter values (Param., row) of the 200 Virgo halos and of the selected random halos. Cluster-size random halos are selected according to the set parameter at redshift 0 (Param., column). A long dash (frame) stands for an (a highly) atypical property value (according to equation \ref{eq:1}, absolute values above 3 and 9 in standard error units respectively) for the Virgo halos with respect to selected random halos. }
\label{ATbl:3}
\end{table}

\begin{table}
\begin{center} 
\begin{tabular}{lcccccccc}
\hline
\hline
&\multicolumn{8}{c}{X$_0^{\rm s}$: additional set parameter distribution at redshift 0}\\
& (1) & (2) & (3) & (4) & (5) & (6) & (7) & (8) \\
Param.  & M$_{200}$ & R$_{200}$ & $\sigma_v$ & N$_{\rm substr}$ & v & CoM-off & C$_{\rm NFW}$ & spin \\
\hline
\hline
M$_{200}$          &    /       &             & --         &  --         & \fbox{ } & \fbox{ } & \fbox{  } & \fbox{ }  \\
R$_{200}$           & --        & /           & --         &  --         & --         & \fbox{ } & \fbox{ } & \fbox{ }  \\

$\sigma_v$         & --         & --          &  /          &     --    &    --    & \fbox{ } & --         & \fbox{ } \\
N$_{\rm substr}$ &            &             &            &     /       & --         &    --      &   --        & --  \\

v                          & \fbox{ } & \fbox{ } & \fbox{ } & \fbox{ } &   /     & \fbox{ } & \fbox{ } & \fbox{ }\\
CoM-off               & --        &  --         & --          &    --       & --       & /         &  --          &   --\\
C$_{\rm NFW}$  & --        & --          & --          &     --       &  --     &  --       &  /           &  --  \\
spin                     & --        & --          &\fbox{ }          &  --& --      &  --       &  --         &    /   \\
\hline
\hline
\end{tabular}
\end{center}
\vspace{-0.25cm}
\caption{Ratio of the standard deviations  (Param., row) of the parameters of the 200 Virgo halos and of the selected random halos. Cluster-size random halos are selected according to the set parameter at redshift 0 (Param., column).  A long dash (frame) stands for a property that is (highly) constrained for the Virgo halos with respect to the selected random halos (values below 0.95 and 0.55 respectively).}
 \label{ATbl:4}
\end{table}


\bibliographystyle{mnras}

\bibliography{biblicomplete}

 \label{lastpage}
\end{document}